\documentclass[12pt]{spieman}  
\usepackage{amsmath,amsfonts,amssymb}
\usepackage{graphicx}
\usepackage{setspace}
\usepackage{tocloft}

\usepackage{amssymb}
\usepackage{comment}
\usepackage{subcaption}
\usepackage{wrapfig}
\usepackage{longtable}
\usepackage{url}
\usepackage{hanging}
\usepackage{hyperref}
\usepackage{multicol}
\usepackage{multirow}
\usepackage{overpic}
\usepackage[inline]{enumitem}
\usepackage{setspace}
\usepackage{comment}
\usepackage{multicol}
\usepackage[normalem]{ulem}
\usepackage[table,xcdraw,dvipsnames]{xcolor}
\graphicspath{{figures/}}

\newcommand\red[1]{\color{black}#1 \color{black}} 

\title{WaveDriver: a Laser Guide Star AO System for HWO}

\author[a]{Benjamin L. Gerard}
\author[a]{Alex Geringer-Sameth}
\author[b]{Aditya R.\ Sengupta}
\author[a]{Alexx Perloff}
\author[a]{Dominic F. Sanchez}
\author[a]{Peter Waswa}
\author[a]{Cesar Laguna}
\author[b]{Rebecca Jensen-Clem}
\author[a]{Lisa Poyneer}
\author[a]{Megan Eckart}

\affil[a]{Lawrence Livermore National Laboratory}
\affil[b]{University of California Santa Cruz}

\cftpagenumbersoff{figure}
\cftpagenumbersoff{table} 
\begin{document}

\maketitle
\begin{abstract}
Habitable Worlds Observatory (HWO) presents a key challenge for technology development in the coming years, requiring a $>$ $100\times$ more stable system than \textit{JWST}. WaveDriver is a concept for a laser guide star spacecraft coupled to an adaptive optics (AO) system onboard HWO that would enable HWO to reach its picometer-level wavefront stability requirements while relaxing other HWO subsystem requirements. At LLNL and UCSC we are revisiting the concept initially proposed by Douglas et al.\ (2019). We present key results key initial results from the first phase of our project, including (1) AO control developments, including with Linear Quadratic Gaussian control and machine learning, (2) AO wavefront sensor (WFS) trade study simulations, and (3) simulations of a photonic lantern natural guide star WFS. A key finding from our work is that WaveDriver could be needed to enable HWO's primary mirror segment stability and/or low order wavefront stability requirements.
\end{abstract}


{\noindent \footnotesize\textbf{*}Benjamin L. Gerard,  \linkable{gerard3@llnl.gov} }

\begin{spacing}{2}   

%
\section{Introduction and WaveDriver Concept}
\label{sec:intro}
The 2020 Decadal Survey of Astronomy and Astrophysics\cite{astro2020} prioritized space-based habitable exoplanet imaging in the coming decades as a top priority, subsequently leading to the ongoing development for NASA's Habitable Worlds Observatory (HWO). However, at the onset of this development it is clear that if HWO were to launch today with \textit{JWST} and/or Roman Space Telescope technogies it would not meet the technology requirements needed to ultimately measure exo-Earth occurence rates\cite{Ganel2025}. One key technology gap related to this is contrast stability, which relates to keeping HWO sufficiently stable such that contrast remains below the required level for such exo-Earth detection and characterization, which can analogously be translated into a HWO system-level wavefront stability requirement of 10 pm rms over 10 minutes per spatial mode, hereafter $\sigma_{10}$.\cite{Douglas2019} 

\red{
HWO's adaptive optics (AO) systems may be used to enable reaching $\sigma_{10}$ if such requirements cannot be met passively, including using of the primary mirror (M1) laser metrology system that actuates the M1 segments and the coronagraph's low and high order wavefront sensors (WFSs) and deformable mirrors (DMs)\cite{Redding2024}. Although some work addressing longer timescale disturbances suggests that AO system requirements relative to $\sigma_{10}$ can be relaxed for some lower order modes (e.g., Refs. \citenum{Pueyo2022, Sanchez-Soria2025}), no such work has been completed at the higher temporal frequencies ($\gtrsim10$ Hz) we consider here, and thus we will assume $\sigma_{10}$ requirements for all modes integrating in temporal frequency space up to the considered AO system's temporal Nyquist limit. In reality, temporal aliasing effects (i.e., where higher temporal frequency wavefront disturbances are seen by other slower WFSs as an unphysical slower wavefront disturbances) are likley dependent on coronagraph type and may allow relaxation of wavefront stability requirements at these higher temporal frequencies for at least some spatial modes. However, we are not aware of any work in this area and will thus assume $\sigma_{10}$ integrated across all relevant temporal frequency ranges to set requirements in this paper.
}

\begin{figure}[h]
    \centering
    \vspace*{-5pt}
    \includegraphics[width=1.0\linewidth]{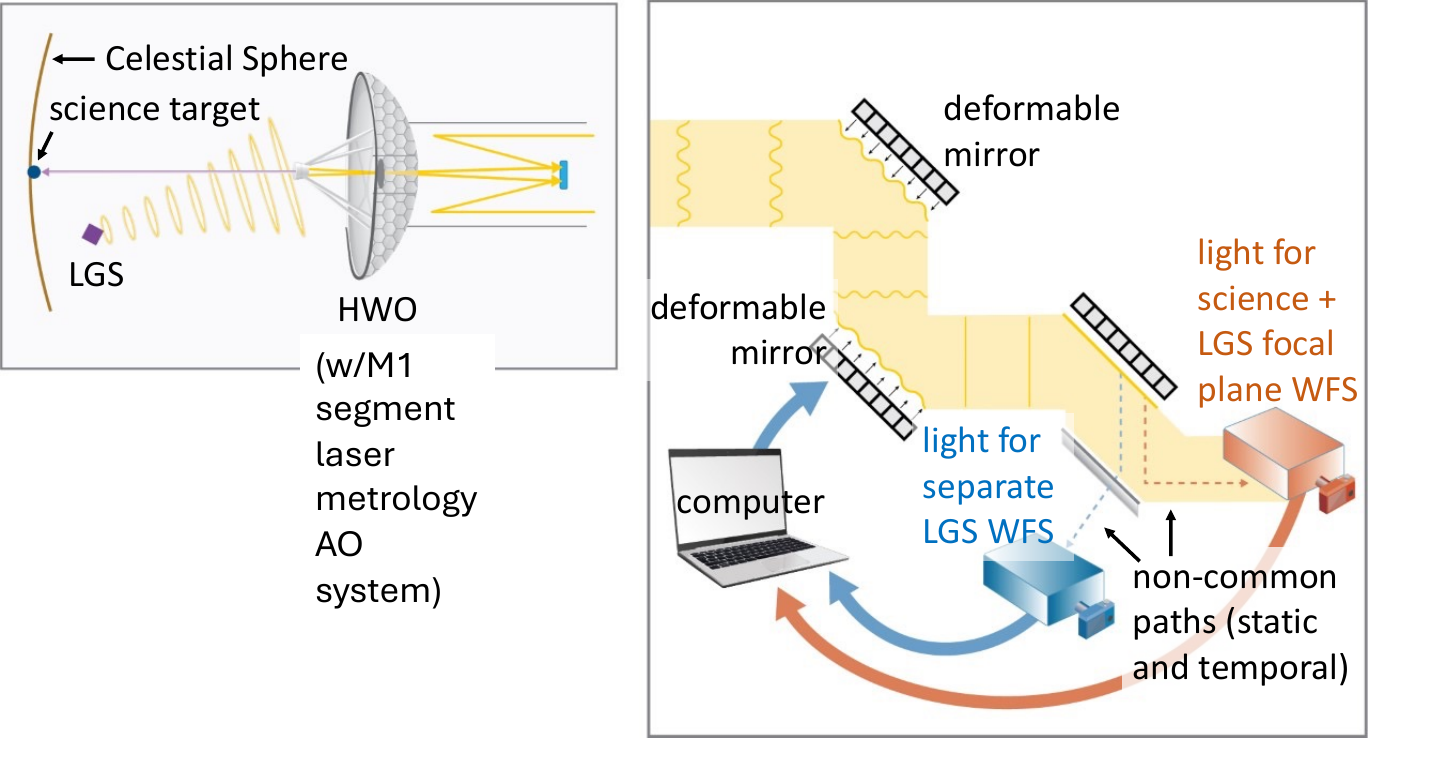}
    \vspace*{-20pt}
    \caption{Conceptual Illustration of the HWO Laser Guide Star (LGS) adaptive optics (AO) system concept considered here, called WaveDriver, designed to enable meeting HWO's \red{nominal} $\sigma_{10}$ wavefront stability requirement.}
    \vspace{-8pt}
    \label{fig:concept}
\end{figure}

Work that led to Refs. \citenum{Douglas2019}, \citenum{Clark2020}, and \citenum{Potier2022} has considered the concept of a laser guide star (LGS) spacecraft to illuminate a Habitable Worlds Observatory (HWO)-like telescope to enable adaptive optics (AO) correction for enhanced telescope+instrument wavefront stability. We revisit and build on these developments here explicitly for HWO, now called ``WaveDriver'' (naming credit: Keith Jahoda) and illustrated in Fig.~\ref{fig:concept}.
Note that the prior work referenced above does not explicitly consider the joint use of HWO's existing internal laser metrology system for primary mirror segment stabilization \cite{Carrier2025}, whereas here, as shown in Fig.~\ref{fig:concept}, we consider this in comparison to WaveDriver. 

\red{
Related to comparing WaveDriver and the HWO M1 laser metrology system, recent work has indicated that HWO's laser metrology system is unlikely to actuate the primary mirror segments faster than 10 Hz due to telescope structural modes expected to be in the 25 to 60 Hz range (Ref. \citenum{Carrier2025} and private communication with J. Tesch). Prior work in Ref. \citenum{Redding2024} considered acquiring HWO laser metrology measurements at up to 100 Hz despite this 10 Hz actuator bandwidth limit in the potential application of open loop control with a higher bandwidth segmented DM (D. Redding, private communication). However, no further work in this area has been published. Furthermore, broader community discussions in 2024 within the HWO Wavefront Sensing and Control Technology Assessment Group led to a de-prioritization in further development segmented DMs. Instead, in this paper, we consider solutions that would use a higher temporal bandwidth DM in closed-loop to overcome this 10 Hz limit. Ref. \citenum{Potier2022} also found similar higher temporal bandwidth DM solutions to this problem; in this paper we build on this work by spanning a larger region of input disturbance and AO system parameter space to help inform potential WaveDriver-illuminated AO system requirements.
}

Here we consider LGS wavefront sensor (WFS) trade study simulations in \S\ref{sec:wfs}, using a photonic lantern (PL) as the natural guide star (NGS) WFS in \S\ref{sec:pl}, and AO Control simulations in \S\ref{sec:control}.
\section{Laser Guide Star Wavefront Sensor Simulations}
\label{sec:wfs}
Ref. \citenum{Douglas2019} considered the use of WaveDriver with a Zernike WFS (ZWFS) for primary mirror (M1) segment stabilization via active M1 segment wavefront control. We revisit this topic here,
\red{
instead using a slightly more sensitive ZWFS with a 2 $\lambda/D$ phase dimple diameter\cite{Chambouleyron2021},
}
by comparing additional WFSs to the ZWFS and also considering the need for correction of lower spatial frequency errors with a separate deformable mirror (DM). The DM-based pupil chopping WFS\cite{Soto2023} \red{(PCWFS)} and a Mach-Zehnder interferometer WFS (MZWFS) were simulated in addition to a ZWFS.
We used 
\red{
configurations similar to
}
HWO exploratory analysis case 1 (EAC1) and 3 (EAC3),\cite{Lou2024} 
\red{
which are off- and on-axis telescope designs that use hexagonal and keystone M1 segments, respectively. We are aware that our EAC1 and 3 models are not exact matches to the total number of segments in NASA's proposed architectures but for the purposes of this exploratory and comparative work differences can be considered negligible. However to avoid confusion subsequent references to each configuration will be noted ``EAC1- and/or EAC3-like.''
}
\red{
Also note that all simulations here assume a central wavelength of 500 nm unless otherwise noted. See appendix \ref{sec:methods_ho} for further details about the configurations and methods used in this section.
}

\begin{figure}[!h]
    \centering
    \begin{subfigure}[b]{1.0\textwidth}
        \includegraphics[width=\textwidth]{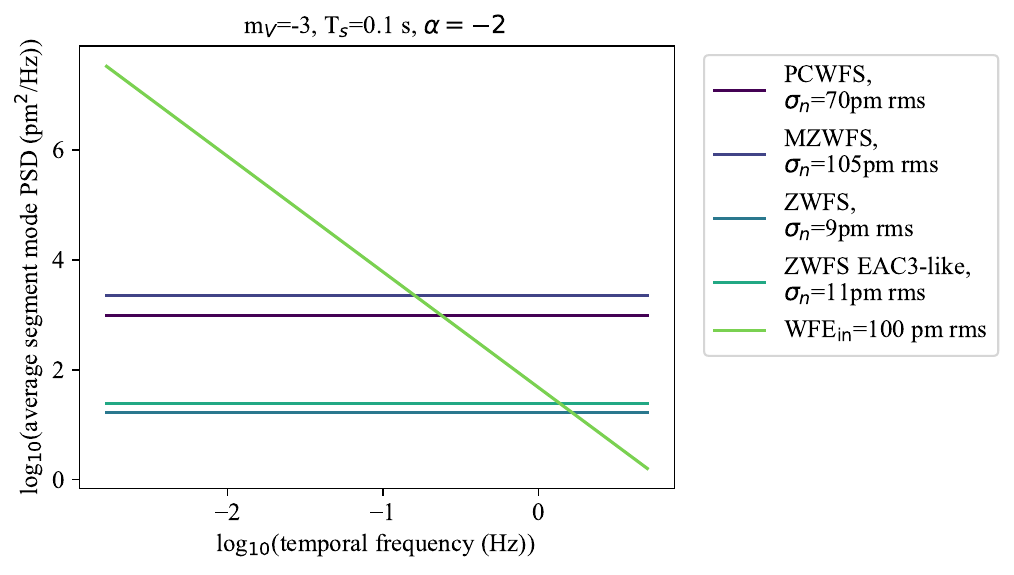}
        \vspace{-20pt}
    \end{subfigure}
    \vspace{5pt}
    \caption{LGS WFS trade study results, showing (1) a Zernike WFS (ZWFS) is the most sensitive WFS considered here (with negligible differences between EAC1 and EAC3
    \red{
    -like cases
    }
    ) and (2) a LGS ($\text{m}_{\text{V}}\lesssim-3$) is needed to reach HWO's $\sigma_{10}$ requirements for M1 segment stability. All curves unless noted are for our EAC1
        \red{
        -like simulation
        }
        .
        }
    \label{fig:lgs_wfs}
\end{figure}

The nominal HWO requirement considered here is $\sigma_n < \sigma_{10}$. Bode's Theorm (i.e., balancing low frequency rejection with high frequency overshoot) implies that open loop photon noise limits 
\red{
($\sigma_n$; see Appendix \ref{sec:methods_ho})
}
can be thought of as limits to achievable closed loop wavefront error with optimal linear control (see \S\ref{sec:control}) and can thus be considered fundamental limits for the purposes of this analysis that can be compared to the $\sigma_{10}$ requirement. Fig.~\ref{fig:lgs_wfs}
\red{
shows the main results for our LGS WFS analysis in response to segment-mode WFEs.
}.

This same simulation framework is also used to simulate photon noise limits for injected low order error wavefront errors, for which we find with a ``perfectly achromatic'' Zernike WFS (i.e., flux scaling with V band but with no PSF magnification and/or scalar mask chromaticity, representing an upper limit on achievable sensitivity for this WFS) 
\red{
with a 15 pixels/pupil plate scale
}
on a $\text{m}_{\text{V}}=3$ star (i.e., among the brightest targets for HWO\cite{Tuchow2025}), that with input temporal WFE (WFE$_\mathrm{in}$) set to 100 pm rms per mode, $\sigma_n=$
\red{
25
} pm rms for an average the first 15 Zernike modes. This suggests that a LGS may be needed to reach $\sigma_{10}$ for low order modes, although other more sensitive WFSs such as the PL could potentially 
\red{
increase
}
limits to fainter stars, as work in \S\ref{sec:pl} will suggest. 
\section{Photonic Lantern as a Natural Guide Star Laser Guide Star Wavefront Sensor}
\label{sec:pl}
\begin{figure}[h]
    \centering
        \includegraphics[width=0.9\linewidth]{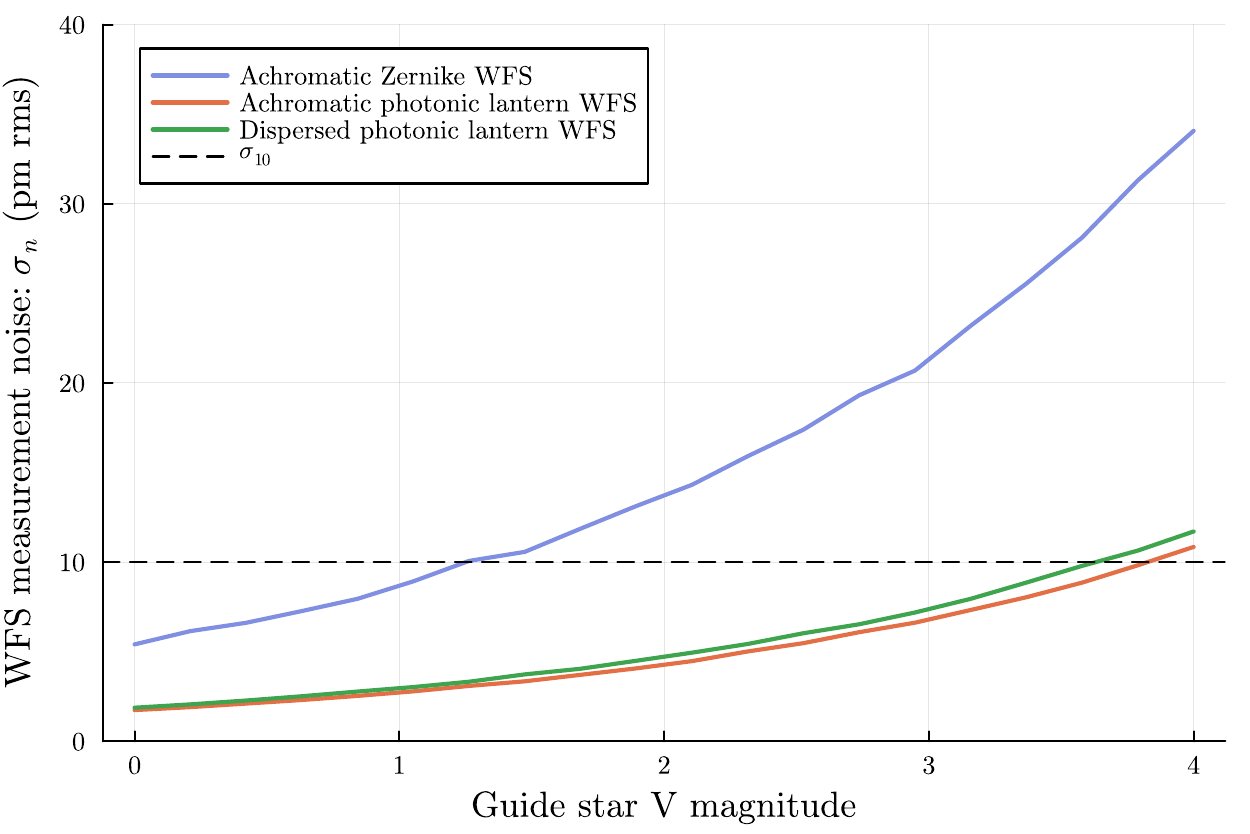}
    \caption{The \red{dispersed} PL meets the $\sigma_{10}$ requirements for tip/tilt/focus for stars with $m_V \sim $
    \red{
    3.6
    }
    or brighter, while analogously a ZWFS requires stars brighter than m$_V\sim 1.3$.}
    \label{fig:measurement_noise}
\end{figure}

In the context of Fig.~\ref{fig:concept}, we consider a photonic lantern (PL\cite{Birks2015}) as a NGS WFS in WaveDriver, motivated by the need for, at minimum, NGS tip/tilt/focus measurements 
\red{
across the full pupil aperture (i.e., not across individual M1 segments)
}
that would be discrepant with equivalent LGS WFS measurements due to WaveDriver spacecraft motion. 
\red{
A PL is a fiber waveguide device that converts multiple wavefront modes of starlight that is injected into one end (called the ``multi-mode end'') into a bundle of separately guided single modes on the other end (called the ``single mode end''). For the purposes of this paper, we consider the measurement noise from photon noise for a PL (which thus far has not been quantitatively simulated by prior work) in response to the lowest order of such modes: tip, tilt, and focus. Leveraging the same methods used in \S\ref{sec:wfs} and described in Appendix \ref{sec:methods_ho} to generate WFS measurement noise, $\sigma_n$, in comparison to $\sigma_{10}$, 
}
we simulate the effect of measurement noise on wavefront reconstruction using a PL for tip/tilt/focus in V band, to ensure closed-loop performance when using a PL as an NGS WFS meets the 
\red{
$\sigma_{10}$
}
HWO requirement. We simulate a 4-port PL 
\red{
in both an ``achromatic'' upper limit and a more realistic spectrally dispersed mode, both described further in Appendix \ref{sec:methods_lo}. We compare PL performance to a 3 pixels/pupil plate scale achromatic V band ZWFS, using the same achromatic properties as used in \S\ref{sec:wfs} to provide an ``upper limit'' of possible ZWFS sensitivity. Fig. \ref{fig:measurement_noise} shows the $\sigma_n$ vs. V band guide star magnitude for the achromatic ZWFS vs. our 4 port PL. Also see Appendix \ref{sec:methods_lo} for additional comparative ZWFS vs. PL analysis and discussion.
}

Figure~\ref{fig:measurement_noise} shows how $\sigma_n$ for tip/tilt/focus scales as a function of the limiting magnitude in comparison between a PL and achromatic ZWFS in V band. A key result from this analysis is that the PL seems to be more sensitive, with respect to photon noise, than a ZWFS for tip/tilt/focus, opening up more accessible targets for HWO to observe, with or without WaveDriver. Note that although this result may seem discrepant with past WFS sensitivity analyses (e.g., Ref \citenum{Chambouleyron23} and \citenum{Guyon2005}), such analyses, which parametrize a unitless variable, $\beta$, do not necessarily isolate WFS measurement noise from an AO error budget perspective and are not analogous to the methods we present here. A PL is also not considered in any such previous WFS sensitivity analyses thus providing no reference for a relative comparison, thus here we provide a first insight into how sensitive PLs seem to be.

\red{
However, a more fundamental point illustrated by Fig. \ref{fig:measurement_noise} is that even the most sensitive possible WFSs fundamentally cannot enable $\sigma_{10}$ for tip/tilt/focus modes at up to a 5 Hz Nyquist limit for stars brighter than m$_V\sim4$. Instead, for a m$_V\sim9$ star (among the dimmest magnitudes HWO would observe\cite{Tuchow2025}), $\sigma_n$ for an achromatic PL reaches 108 pm rms for the average of tip/tilt/focus modes when integrated out to 5 Hz but instead reaches 10 pm rms when integrated out to (10$^2$ / 108$^2$) $\times$ 5 = 0.04 Hz, which as a Nyquist limit would correspond to 0.08 Hz being fastest possible tip/tilt/focus loop rate that could operate on such a star. Conventional AO control theory then sets the highest 0 dB bandwidth at 10 times below this limiting frame rate: 0.008 Hz. This analogously in principle requires that temporal disturbances at all frequencies above 0.008 Hz (i.e., timescales shorter than about 2 minutes) must be passively stable below 10 pm rms, in principle with no active wavefront correction possible to relax such stability requirements (unless any internal and/or external laser had $\ll\sigma_{10}$-levels of temporal non-common path error with respect to the `truth'' NGS tip+tilt+focus errors).
}
\section{Adaptive Optics Control Simulations}
\label{sec:control}
In this section we build on \S\ref{sec:wfs} and \S\ref{sec:pl}, which considered WFS measurement noise, to  consider possible temporal bandwidth error limitations to reaching $\sigma_{10}$ (\S\ref{sec:ai} and \ref{sec:fpga}) and combine bandwidth and measurement error into a single analysis (\S\ref{sec:ao_bandwidth_error}).
%
%
\subsection{Reinforcement Learning}
\label{sec:ai}
We developed several AI-based controllers using reinforcement learning (RL) and assuming the use of an FPGA. As will be motivated in \S\ref{sec:fpga}, we determined that realistic computational latency for M1 segment mode wavefront control calculations using a circular buffer with the past 100 frames per mode as an input into the RL controller at each new WFS frame would be of order a few $\mu$s. We adopt a total AO system latency of 1.5 ms based on typical performance (including WFS camera read out, WFS camera frame transfer, computation, DM frame transfer, and DM actuation), for which bandwidth error is dominated by the 100~ms integration time that represents the maximum possible HWO M1 segment actuation speed.\cite{Carrier2025}

The AI framework used neural networks as controllers. We tested multi-layer perceptrons (MLP) for these networks with two or three hidden layers of width 128, 256, or 1024. Since MLPs do not have an internal memory, we incorporated the time series history by having a controller take as input the current wavefront sensor measurement along with the previous 100 WFS measurements and previous 100 control outputs. The controller then returned a scalar control value for a single mode of the deformable mirror.

We trained the controllers in a time-domain reinforcement learning environment we created. Several machine learning modalities were used for training, including supervised learning on ``perfect predictive'' control and reinforcement learning using a deep deterministic policy gradient (DDPG) algorithm inspired by Ref. \citenum{Nousiainen22}. In supervised learning we train the controller by looking ahead in the time series and calculating a control value that would equal the average wavefront error over the duration of the zero-order hold. We call this perfect prediction. The neural network controller is then optimized by gradient descent to reproduce this perfect control value. In DDPG learning, we train the controller network simultaneously with a dynamics model. The dynamics model is a second neural network that predicts the next WFS measurement given the current observed state plus a trial control value. The controller network is trained by feeding it through the dynamics model and using gradient descent to minimize the absolute value of the predicted next WFS measurement. Unlike training on perfect prediction, DDPG (or other reinforcement learning) can be trained on a physical system -- not just in simulation -- as the only training data are WFS measurements, not the true, but unobservable, wavefront error.

\begin{figure}[h]
    \centering
    \includegraphics[width=1.0\linewidth]{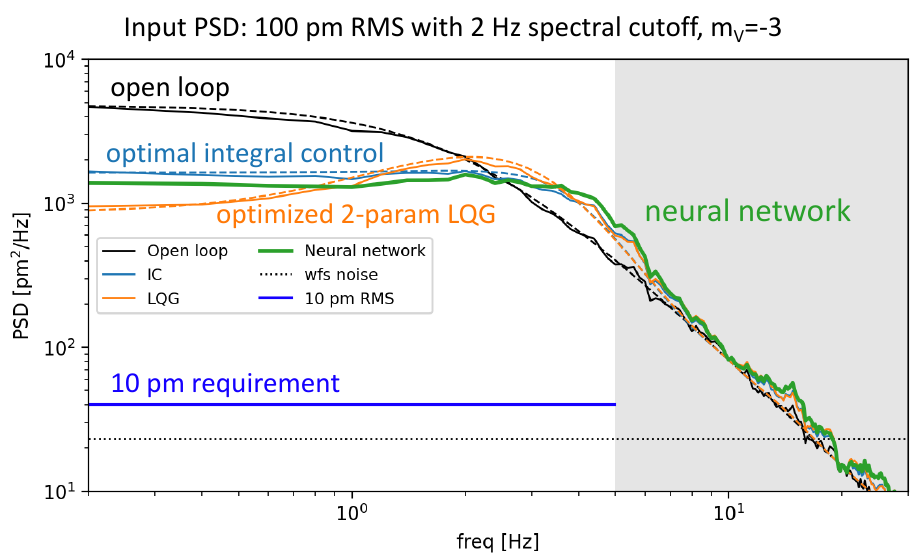}
    \vspace*{-10pt}
    \caption{AO bandwidth error may prevent the HWO M1 laser metrology wavefront control system, which has a 10~Hz actuation limit\cite{Carrier2025}, from reaching HWO's 10 pm requirement ($\sigma_{10}$). The input ``open loop'' PSD reflects integrated modeling results from Ref. \citenum{Potier2022}, with various closed-loop PSDs (light blue, green, orange) corresponding to different controllers clearly not reaching $\sigma_{10}$ (dark blue), despite the measurement noise (dotted black) being below $\sigma_{10}$.}
    \label{fig:control}
\end{figure}

We benchmarked performance by comparing the AI controllers to optimal leaky integral control\cite{Gendron94} and Linear Quadratic Gaussian (LQG) auto-regressive order 2 controllers using the framework from Ref. \citenum{Poyneer2023}. We found that the neural network controllers achieved the same performance as optimal integral control and LQG. The results of this tuning and control optimization are shown in Fig.~\ref{fig:control} for controller trained with supervised learning (reinforcement learning training was similar).
The shaded region in Fig.~\ref{fig:control}
\red{
(produced by including subframes within a 10 Hz loop rate simulation with matching temporal statistics to the sub-Nyquist frequencies)
}
is above the 5 Hz Nyquist limit of the M1 laser metrology wavefront control system\cite{Carrier2025}, but a higher temporal bandwidth high order DM and/or segmented DM (i.e., illuminated by WaveDriver) could potentially reduce this bandwidth error below $\sigma_{10}$, to be discussed further in \S\ref{sec:ao_bandwidth_error}.  A more exhaustive search of hyperparameters along with more advanced neural network architectures for the controller and dynamics models
\red{
would help determine to what extent, if any, this proposed AI AO control architecture could enable reaching $\sigma_{10}$ in cases like those in Fig. \ref{fig:control}. More specifically, it is particularly interesting to consider cases where it is impossible to reach $\sigma_{10}$ via optimal linear control (i.e., it would violate Bode's theorem) but may be possible via non-linear AI-based control (e.g., if it is possible to have a controller whose bandwidth it at or above the frame rate). Thus, further work in this area
}
may be considered in the future.
\subsection{FPGAs}
\label{sec:fpga}
%

Field-programmable gate arrays (FPGAs) are highly useful for low-latency, power-constrained applications where the problem fits a dataflow model and one can invest in a hardware-oriented implementation. Where Graphical Processing Units (GPUs) rely on Single Instruction, Multiple Data (SIMD) lanes, executing the same instruction over different data, FPGAs can efficiently exploit task-level, pipeline, and spatial parallelism, wiring different kernels side-by-side. Additional efficiency can be gained by performing a bit-level optimization of numeric formats. We wanted to see if the neural networks used in~\S\ref{sec:ai} could be implemented on an FPGA and if the latency would be low enough to justify the added development costs and complexities of using an FPGA.

Over the past decade, much development has gone into the implementation of ML algorithms on FPGAs. Companies like AMD (formerly Xilinx) and Intel (formerly Altera) have developed tools for commercial industries. However, until recently, those tools targeted applications requiring microsecond latencies and moderate throughput for ML inference tasks. The LHC experiments at CERN were looking to utilize ML and FPGAs for their own purposes, but needed much lower latencies and higher throughputs. Tools such as hls4ml~\cite{fastml_hls4ml, Duarte:2018ite, Fahim:2021cic} and Conifer~\cite{Summers_2020} were developed to allow for this, even being used with more complicated convolutional or recurrent neutral network architectures~\cite{Aarrestad_2021, Khoda_2023}. 
Follow-on tools such as rule4ml~\cite{IMPETUSUdeSrule4ml} and wa-hls4ml/lui-gnn~\cite{10.1145/3706628.3708827, 2824692} are AI systems trained on thousands of synthesized NN models turned into FPGA kernels. These tools can be used to estimate the resulting FPGA resource utilization and latency without the need to go through a time-intensive hyperparameter tuning process.


We started by utilizing one of the NN architectures of~\S \ref{sec:ai} with 256 input nodes, two hidden 256 node hidden layers, and an output layer with one node. Each of these layers utilized a Rectified Linear Unit activation function. 
In order to accurately estimate whether or not our firmware would meet the resource utilization and latency requirements, we needed to define a target platform. We chose to study implementation on an AMD Alveo U200 card with an UltraScale+ XCU200-2FSGD2104E FPGA~\cite{AMD-DS962, AMD-UG1289}. Although this may not be the FPGA utilized for a space-based application, it was an appropriate analog for ground-based testing. Table~\ref{tab:XCU200_resources} shows the resources available on this FPGA. 

\begin{table}[t]
    \centering
    \begin{tabular}{c|c}
         \textbf{Resource} & Count (K) \\\hline
         Look-up tables (LUTs) & 1,182 \\
         Registers & 2,364 \\
         Digital Signal Processing (DSP) slices & 6.84\\
         Block RAMs (36kb) & 2160 \\
    \end{tabular}
    \caption{The programmable logic resources available on an AMD UltraScale+ XCU200 FPGA. The listed resources are the total amounts on the chip. The usable amount within the dynamic region of the programmable logic are lower due to the Deployment Shell Architecture implemented on the static region of the programmable logic. For example, the dynamic region contains 1766 Block RAMs as 394 RAMB36 primitives are used for the PCIe and DDR4 interfaces.}
    \label{tab:XCU200_resources}
\end{table}

Rather than using the AMD Vivado/Vitis HLS~\cite{AMD-UG902, AMD-UG1399} backends and hls4ml directly, we chose to utilize rule4ml to speed-up the resource estimate workflow. However, the tool was unable to provide accurate resource estimates for the chosen NN. Instead, we used a smaller NN with 111 nodes in each layer. For the chosen platform, with the $\textrm{ap}\_\textrm{fixed}<16, 6>$ precision, and utilizing each DSP only once (minimal reuse of resources), the estimated latency was 266.48 clock cycles ($\sim$2.66 ${\mu}s$ with a 10 $ns$ clock) utilizing 2.01\% of the available Block RAMs, 3.39\% of the DSPs, and 46.46\% of the LUTs.
%
This small feasibility study shows that it is possible to implement a moderately sized NN on an FPGA for use within a low-latency AO system.
\subsection{AO bandwidth error limitations}
\label{sec:ao_bandwidth_error}
\begin{figure}[h]
    \centering
    \includegraphics[width=1.0\linewidth]{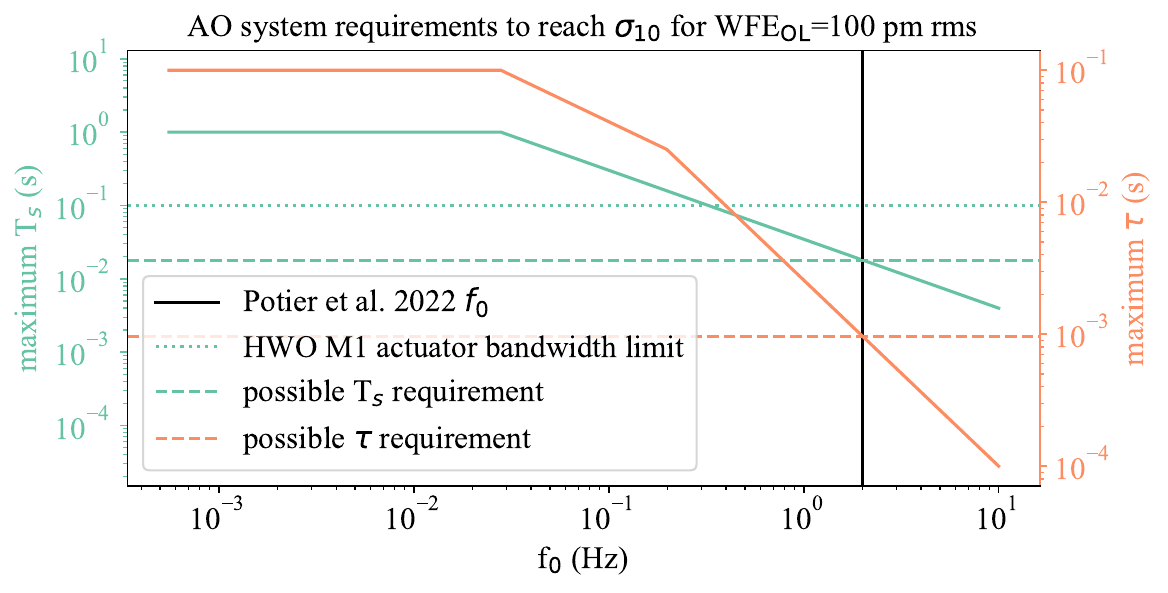}
    \vspace*{-20pt}
    \caption{\red{The 10 Hz HWO M1 actuator temporal bandwidth limit\cite{Carrier2025} may prevent HWO from reaching $\sigma_{10}$. This figure shows the required AO system parameters---frame rate ($T_s$; solid teal line) and latency ($\tau$; solid orange line)---needed needed to reach $\sigma_{10}$ as a function of the input temporal disturbance spectral break ($f_0$) when the integrated rms WFE stability between 1/(10 minutes) and the WFS' temporal Nyquist limit (WFE$_\mathrm{in}$) is 100 pm rms as in Ref. \citenum{Potier2022}. The left y-axis shows the slowest possible AO system frame rate that can still enable reaching $\sigma_{10}$, and the right y-axis shows the corresponding highest possible AO system latency. Using the spectral break of 2 Hz from Ref. \citenum{Potier2022} (vertical black line), the 10 Hz M1 segment bandwidth limit is too slow to reach $\sigma_{10}$ (i.e., the dotted teal line is above the dashed teal line).}}
    \label{fig:limits}
\end{figure}

To assess the parameter space for potential solutions to bandwidth error and measurement noise limitations to reaching $\sigma_{10}$, we developed and open-sourced the following code in a graphical user interface (GUI) format: \url{https://github.com/LLNL/hwo1dGUI}. \red{We encourage readers to use and explore this GUI, which can by run in a web browser via the mybinder.org link within the github readme file. See Appendix \ref{sec:gui_tutorial} for a tutorial and further guidance of how to use this GUI.} This code was ultimately used to produce Fig.~\ref{fig:limits}, which was made as follows: for a given value of input temporal PSD spectral break (f$_0$), other parameters were selected such that the integrated closed loop WFE (``output WFE'' in the title of the GUI) was less than 10 pm rms by 
\red{(1) first selecting an integrated input temporal WFE (WFE$_\mathrm{OL}$, equivalent to $\sigma_s$ in \S\ref{sec:wfs}) level of 100 pm rms as in Ref. \citenum{Potier2022},} 
(2) choosing the longest possible WFS exposure time (T$_S$; i.e., the slowest possible AO system), then (2) for this chosen T$_S$ then choosing the largest possible AO system latency ($\tau$). \red{All other variables are allowed to vary freely, including input PSD slope ($\alpha$; larger $\alpha$ redistributes input broadband temporal disturbances to higher temporal frequencies and is thus more constraining on other AO system parameters), WFS sensitivity ($\beta$) as defined in Ref. \citenum{Douglas2019}, and maximum V band magnitude (m$_V$) using the same throughput and flux zero point scaling as in Ref. \citenum{Douglas2019}.
} 

Exploring parameters in the GUI and the trends in Fig.~\ref{fig:limits} both illustrate the following key findings of when/how $\sigma_{10}$ can be achieved:
\begin{enumerate}
    \item $\text{m}_{\text{V}}\lesssim -3$. This corroborates the conclusions from Refs. \citenum{Douglas2019}, \citenum{Potier2022}, and \citenum{Carrier2025} that a bright laser source is needed to stabilize the segments, with science target stars being fundamentally too faint to reach $\sigma_{10}$, leaving WaveDriver (external) and laser metrology (internal) as possible LGS solutions.
    \item the input WFE is ``passively'' stable below 10~pm rms. \red{Assuming $\sigma_{10}$ must be met by integrating the noise PSD from 1/10 minutes up to the AO system's Nyquist limit, }this can be equivalently re-stated as requiring passive stability of 8 pm rms between 1 and 5 Hz, where this temporal frequency range is above the 
    \red{
    maximum
    }
    accessible M1 segment laser metrology temporal bandwidth\cite{Carrier2025}, thus requiring passive stability at these high frequencies not accessible to attenuation via active control. Recent HWO EAC1 integrated modeling where micro-thrusters were used instead of gyroscopes\cite{Feinberg2025, Zeimer2025} suggests that this is achievable and is the baseline strategy for reaching $\sigma_{10}$ at \red{$\gtrsim$ 1 Hz }frequencies \red{(but see further discussion in \S\ref{sec:conclusion}).}
    \item integrated input disturbances are above 10 pm rms but the spectral break ($f_0$) and/or power law ($\alpha$) are adjusted to shift power to lower temporal frequencies that are below the $\sim$1 Hz M1 segment temporal bandwidth limit. For example, for $\sigma_s\equiv\text{WFE}_\mathrm{in}=100$ pm rms, $\alpha=-2$, and $f_0<0.25$ Hz, $\sigma_{10}$ can be reached with $T_s=0.1$ s, $\tau=1.5$ ms. This indicates an area potentially worth investigating where system-level observatory design optimization could shift input high frequency input disturbances to lower frequency output disturbances that are within the AO system temporal bandwidth to enable reaching $\sigma_{10}$.
    \red{
    Such optimization is essentially a design mounting problem of how to shift high temporal frequency disturbances to lower temporal frequencies before reaching the optics. This type of optics mounting optimization procedure has been demonstrated with auto-differentiable codes designed for finite element analysis, e.g., Ref. \citenum{Kihm2013}.
    }
    \item\label{pt:wavedriver} using a ``second stage'' high temporal bandwidth DM, which would require WaveDriver. For the 100 pm rms input case from Ref. \citenum{Potier2022} (additionally with $f_0=1$ Hz and $\alpha=-2$), $\sigma_{10}$ can be achieved with $\tau=1.5$ ms for $T_s \lesssim 20$ ms, which such a high temporal bandwidth DM could enable. Although HWO-relevant DM actuation speeds have thus far been limited to speeds $\lesssim$10~Hz to meet voltage stability requirements (e.g., Refs. \citenum{Bendek2020, Bendek2024}), development to enable $\gtrsim$100 Hz actuation speeds while still meeting voltage stability requirements is potentially feasible via DM electronics RC circuit low pass filter cutoff adjustments (G. Ruane, E. Bendek, T. Groff, private communication) and thus worth investigating. Also note that in principle the inherent closed-loop nature of a WaveDriver AO system would relax such high temporal bandwidth DM votage stability requirements at temporal frequencies below the expected WaveDriver temporal bandwidth ($f\lesssim9$ Hz for the example given at the beginning of this bullet point). Also note that it has sometimes been stated that laser metrology ``feed-forward control'' at $>$ 100 Hz frame rates could be enabled with a high bandwidth continuous and/or segmented DM, but this nomenclature is misleading as in reality this would be inherently open loop control, which would (1) add risks of calibration errors and (2) prevent the typical 20 dB/decade of rejection at low temporal frequency (i.e., a controller with gain = 1 and leak = 0, equivalent to open loop control, does not provide such rejection) and thus add additional constraints on other parameters such as maximum input WFE, latency, frame rate, etc.
\end{enumerate}
\section{Conclusion and Discussion}
\label{sec:conclusion}
Wavefront stability such that contrast is maintained at 10$^{-10}$ levels over a science exposure is a key technical requirement for HWO and potentially not currently feasible with current state-of-the art technologies. In this work we build off of previous work\cite{Douglas2019, Potier2022} to propose and motivate WaveDriver, a LGS AO system for HWO, as a solution to bridge this contrast stability technology development gap. Note that more specific WaveDriver mission concept development is beyond the scope of this paper and will be considered in future phases of our work. We first considered different WaveDriver LGS WFS options in \S\ref{sec:wfs}. We determined that a Zernike WFS (ZWFS) produced the lowest measurement noise in comparson to 
\red{
a pupil chopping WFS and Mach-Zehnder interferometer WFS. We also determined that for AO-enabled wavefront stabilization down to $\sigma_{10}$ levels at up to 10 Hz loop speeds,
}
a LGS is needed for M1 segment modes and also potentially needed for lower order Zernike modes 
\red{
across the full M1 pupil, corroborating findings from previous work\cite{Douglas2019,Potier2022}.
}
In \S\ref{sec:pl} we proposed a photonic lantern (PL) as a 
\red{
full pupil tip/tilt/focus
}
NGS WFS in the WaveDriver system and showed that measurement noise for this WFS 
\red{
is lower than a ZWFS but still would not
}
in principle enable HWO to observe most of it's planned targets
\red{
with a 10 Hz loop speed, instead requiring at most a 0.04 Hz loop speed with 10 pm rms passive stability needed above 0.008 Hz.
}
Lastly in \S\ref{sec:control} we investigated potential machine learning and FPGA-based AO control options for WaveDriver and in doing so found a potential $\sigma_{10}$ temporal bandwidth error limitation from M1 segment and/or low order wavefront errors. To address this potential limitation, we proposed potential solution requiring WaveDriver that would use a high temporal bandwidth DM
\red{
to enable DM actuation at $\gtrsim$ 100 Hz.
}
We found such potential temporal bandwidth errors in part by developing and open-sourcing a HWO AO control GUI (\url{https://github.com/LLNL/hwo1dGUI}\red{; see Appendix \ref{sec:gui_tutorial} for a tutorial on how to use this GUI}) that considers such limitations under various input and AO system parameters. 

Expanding further on point \ref{pt:wavedriver} in \S\ref{sec:ao_bandwidth_error}, we are aware that the baseline HWO strategy for high temporal frequency M1 segment stabilization will be to engineer the telescope to meet $\sigma_{10}$ requirements ``passively,'' and indeed more recent HWO-relevant work suggests that this may be the case (e.g., Ref. \citenum{Carrier2025}). However, it is worth noting here that even if WaveDriver is considered a backup plan, a risk assessment should ultimately help guide this possible WaveDriver application. For example, using a standard systems engineering risk matrix, if there is a $>$10\% chance of WFE$_\mathrm{in}>8$ pm rms above 1 Hz, not having WaveDriver as a solution to address this would in principle result in at least medium risk to HWO, which should be evaluated against HWO's overall risk tolerance posture. Also note that point \ref{pt:wavedriver} in \S\ref{sec:ao_bandwidth_error} applies to both high and low order errors (the latter of which is unlikely to be sufficiently corrected by M1 segment control alone) and in \S\ref{sec:wfs} we found a $>2.5\times$ WFS measurement noise gap from $\sigma_{10}$ for HWO NGS targets, and thus use of a WaveDriver-enabled HWO LGS AO system for high-speed active correction of low order errors is still of interest for future technology development. We are also aware that WaveDriver is of interest for low temporal frequency absolute phase correction of M1 segment phasing error to increase sky-coverage relative to planned out-of-field NGS sensing (Jurling et al., private communication). Also note that it is known that the $\sigma_{10}$ HWO requirement (i.e., setting the same flat PSD requirement for all spatial modes) used in this paper should be further developed to ultimately provide specific spatio-temporal PSD requirements, which would then help to inform AO system requirements and/or potential bandwidth error limitations. %
\subsection*{Disclosures}
The authors declare that there are no financial interests, commercial affiliations, or other potential conflicts of interest that could have influenced the objectivity of this research or the writing of this paper.
%
\appendix
\section{Methods}
\label{sec:methods}
\subsection{High Order LGS WFS simulations}
\label{sec:methods_ho}
\begin{figure}[h]
    \centering
    \includegraphics[width=1.0\linewidth]{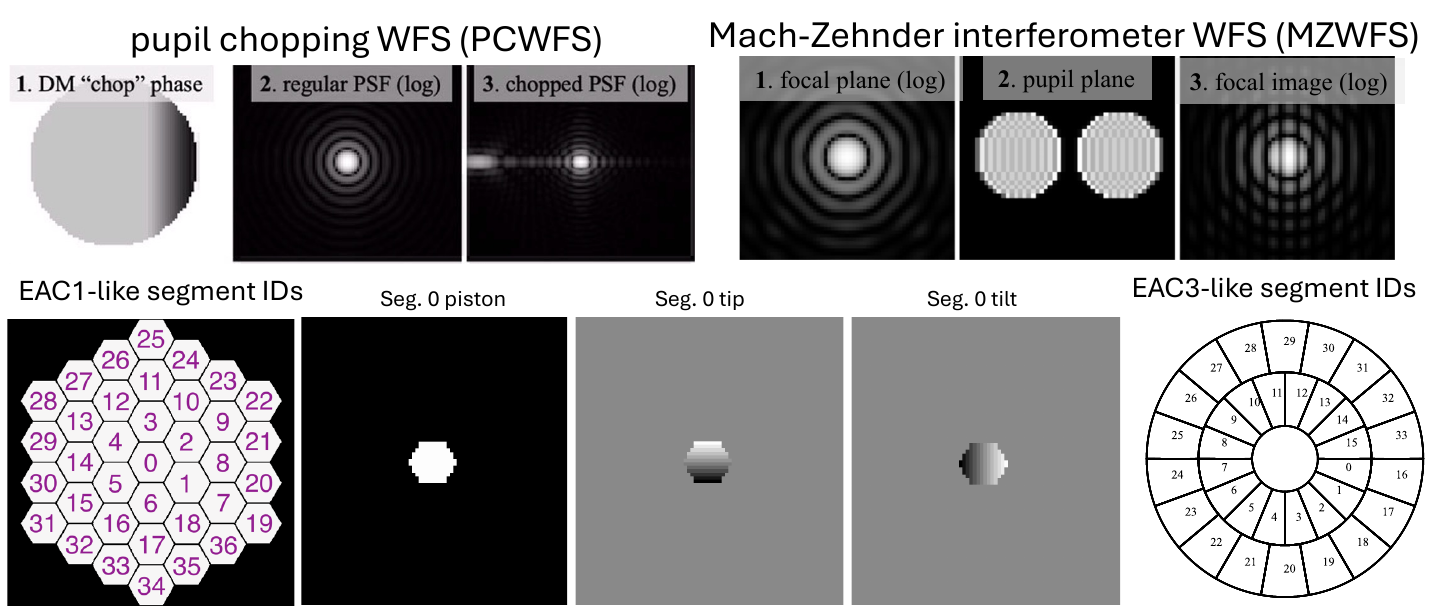}
    \caption{
    \red{
    LGS WFS simulation concept illustrations for the DM-based PCWFS\cite{Soto2023} and MZWFS used in our study in comparison to the ZWFS, and additionally for the EAC1- and 3-like architectures used.
    }
    }
    \label{fig:wfss_and_eacs}
\end{figure}
Illustrated in Fig. \ref{fig:wfss_and_eacs}, our EAC1-like configuration used 37 unobscured M1 hexagonal segments and considered piston+tip+tilt modes for each segment, and our EAC3-like configuration included 34 keystone segments and a 24\% secondary obscuration (but with no spiders for these simulations) for the ZWFS for a comparison of EAC sensitivities. Throughput-related parameters are otherwise matched to Ref. \citenum{Douglas2019}, which were then Fraunhoffer-propagated to form various WFS images. The MZWFS simulates a 50:50 shallow angle focal plane beam-splitter by differentially tilting two realizations of a focal plane electric field such that a downstream pupil plane generates two pupils that are separated by 1.1 pupil diameters, which then interfere in the final focal plane to form fringes. The DM-based PCWFS builds on past work\cite{Gerard2022,Gerard2023,Soto2023} by assuming use of a continuous DM in the chopping WFS path that uses a 0.3 pupil diameter chop fraction (already optimally determined in the above-referenced work on PCWFSs) with 12 radian peak-to-valey (PV) amplitude stroke.
\begin{figure}[h]
    \centering
    \includegraphics[width=0.6\linewidth]{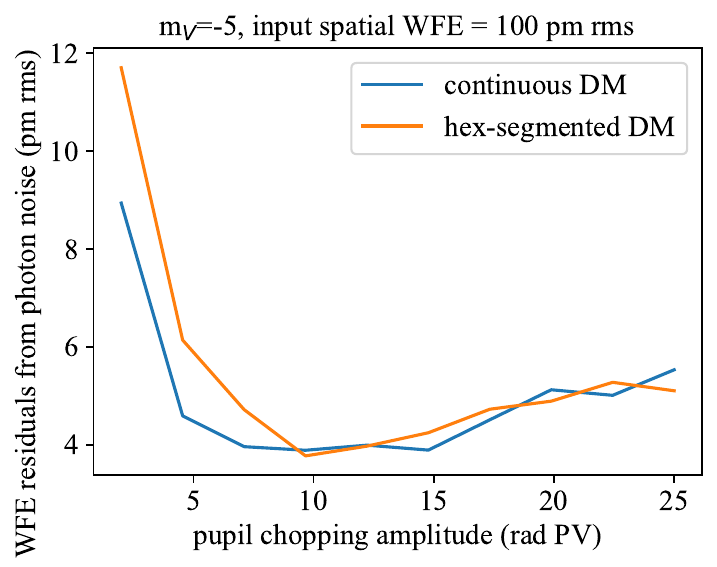}
    \caption{ 
    \red{
    An optimal PCWFS DM chopping amplitude of 12 rad PV produces the lowest amount of WFS measurement noise, computed from the rms of the difference of reconstructed wavefronts from simulated images with and without photon noise for this m$_V$=-5 case. Simulations average 30 random WFE seeds per chop amplitude, each seed with 100 pm rms EAC1-like segment mode phase errors. The two curves, which both use a chop fraction of 0.3 and optimize at a similar location, use a straight DM chop ridgeline (labeled ''continuous DM'') or a ridgeline representing a hexagonal segmented DM at the edge of the pupil whose circumscribed diameter is 30\% of the full pupil diameter.
    }
    }
    \label{fig:pc_amp}
\end{figure}
Fig. \ref{fig:pc_amp} shows how we found a 12 radian PV amplitude to provide optimal PCWFS photon efficiency. 

\begin{figure}[h]
    \centering
    \includegraphics[width=1.0\linewidth]{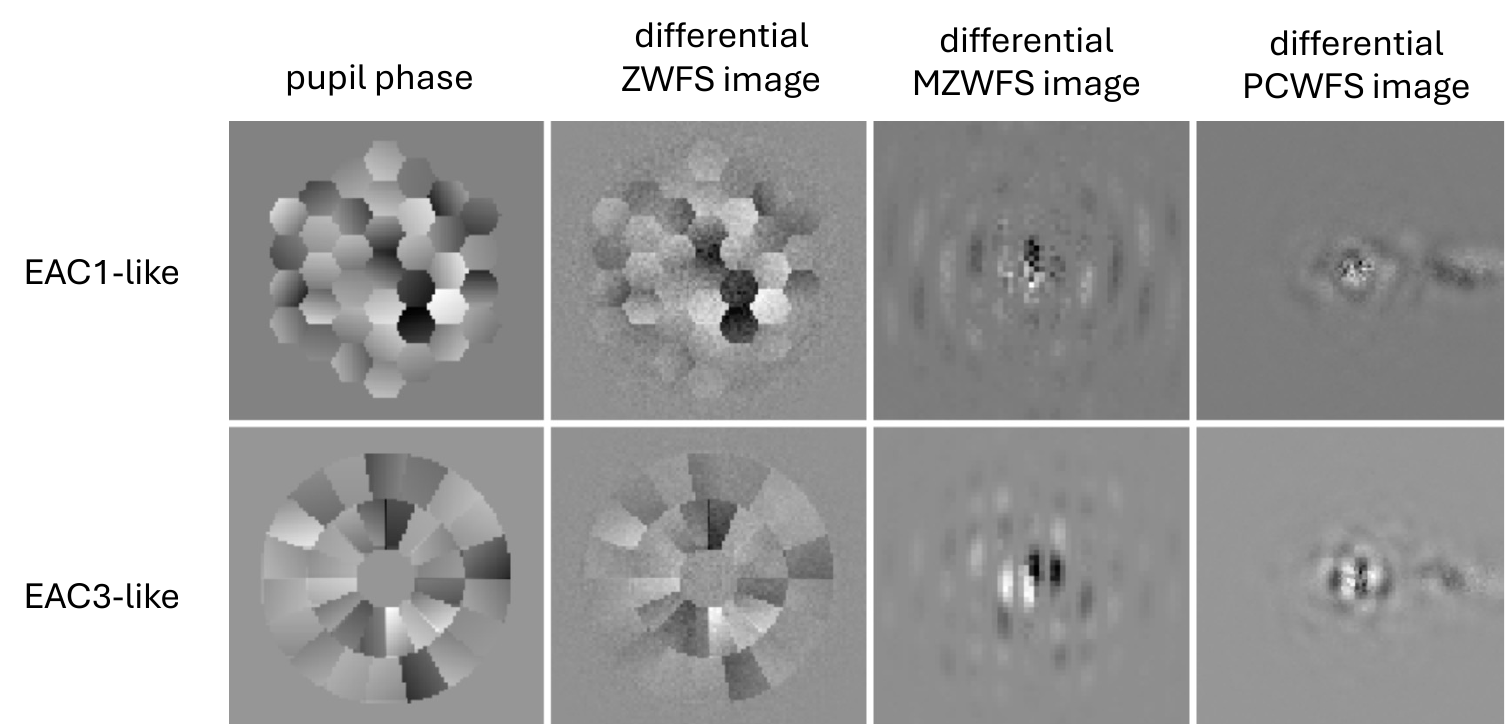}
    \caption{
    \red{
    Simulated phases for EAC1- and EAC3-like cases and corresponding target WFS images (i.e., differential with respect to a noiseless static WFE-only image) for m$_V$=3, spatial WFE = 100 pm rms projected onto the given segment modes.
    }
    }
    \label{fig:spatial_wfe_intro}
\end{figure}

Simulations for all WFSs and EACs include a static wavefront error (WFE) of 1 nm rms, normalized between 0 and 10 cycles/pupil with a -2 power law. All simulations use a standard linear least-squares vector matrix multiply (VMM) reconstructor, where differential images relative to a static image generate an interaction matrix, command matrix via a pseudo-inverse using a singular value decomposition cutoff (SVD) with a normalized SVD cutoff of 10$^{-3}$, ultimately producing modal coefficients via a VMM. WFS linearity via this VMM approach for all simulation cases was verified over the expected range of input disturbances, which is not surprising given the small WFE amplitudes considered here. The simulations inject M1 segment temporal disturbances using random seeded one dimensional time series per mode, with a given power law, $\alpha$, and spectral break, $f_0$, and an input modal temporal WFE, WFE$_\mathrm{in}$, of 100 pm rms, following the ``worst case'' scenario in integrated modeling results from Ref. \citenum{Potier2022}. Photon counting parameters use a Vega magnitude flux zero point \cite{Bessell1998}, a 6m diameter telescope consistent with EAC1 (and also used for EAC3-like simulations to enable an apples-to-apples segment modal comparison), WFS frame exposure time of 0.1s, system throughput of 10\% as in Ref. \citenum{Douglas2019}, and assume use of a WFS camera with no read noise. Fig. \ref{fig:spatial_wfe_intro} illustrates how EAC1- and EAC3-like segment phases are converted into noisy images.

\begin{figure}[h]
    \centering
    \includegraphics[width=0.9\linewidth]{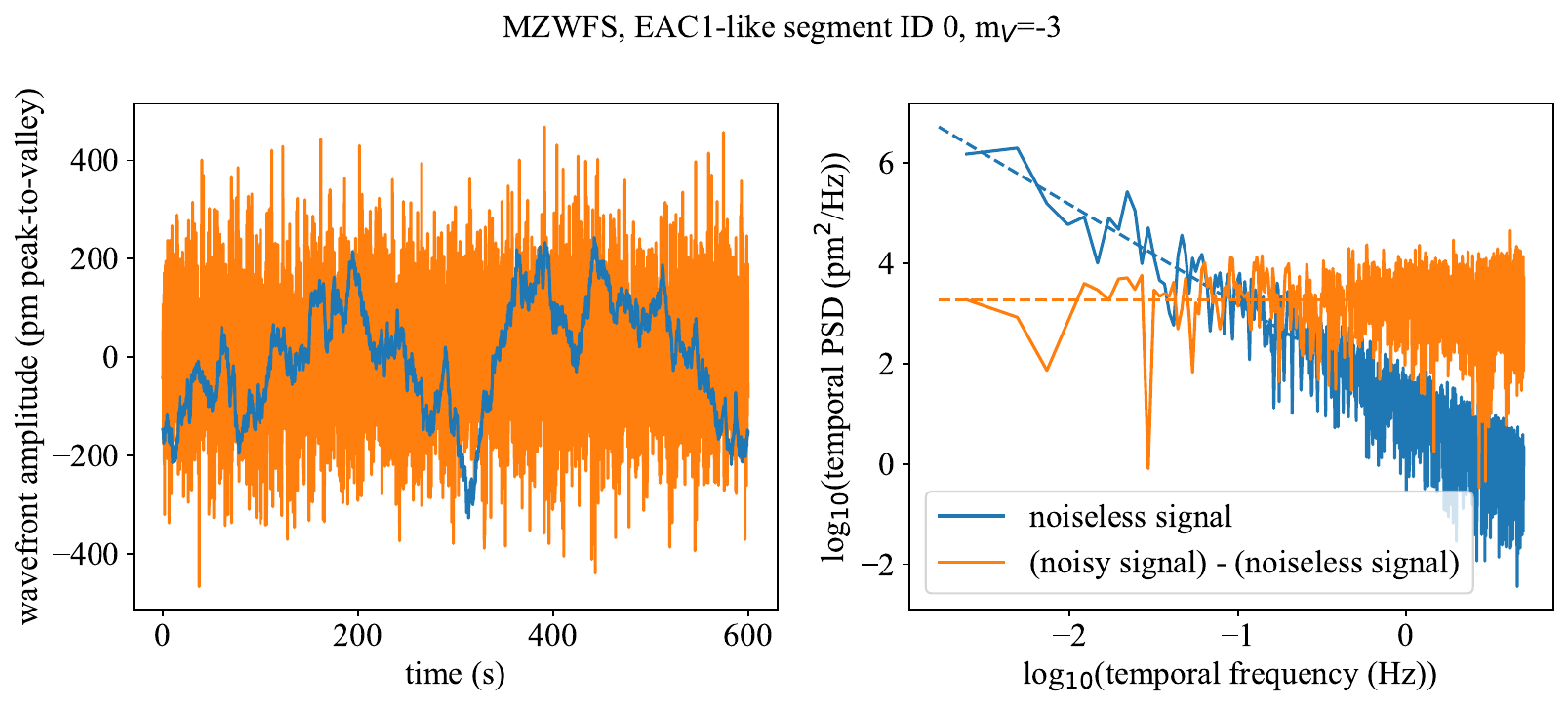}
    \caption{
    \red{
    Reconstructed temporal modal coefficients and corresponding temporal PSDs, for a single segment mode, (1) without photon noise (blue curves) and (2) differential between an image with photon noise and (1). The square root of the integral of the orange PSD in linear space is $\sigma_n$, which in this case is 129 pm rms. Dashed lines in the right panel show a linear fit to the signal and noise components in log-log space. Such fits are what are displayed in Fig. \ref{fig:lgs_wfs}.
    }
    }
    \label{fig:psd_methods}
\end{figure}

The time domain wavefront data cubes are generated at a cadence of 0.1s (with 10 subframes averaged per frame to also simulate temporal aliasing effects up to 100 Hz) and run for 10 minutes in simulations both with and without photon noise. Each phase map at each time stamp is then Fraunhoffer propagated to produce a noisy image like in Fig. \ref{fig:spatial_wfe_intro}. Computing all the reconstructed modal coefficients for a given time series of WFS frames produces a 6000$\times$111$\times$2 array of M1 segment modal coefficients (the last dimension of 2 is for with and without photon noise) for the EAC1-like case and 6000$\times$102$\times$2 for the EAC3-like case. The difference of temporal modal coefficients with vs.\ without photon noise is then used to generate a temporal power spectral density (PSD) curve (using a Hanning window over the full 6000 frames, with no segment overlap due to using just one segment to include the lowest accessible temporal frequency). An illustration of this time domain coefficient reconstruction and corresponding PSD generation to produce $\sigma_n$ is illustrated in Figure \ref{fig:psd_methods}.
Such noise temporal PSDs are then averaged over all M1 segment modes, and this average M1 segment temporal PSD is then integrated over all temporal frequencies between 1/(10 minutes) and the 5 Hz Nyquist limit and then square rooted, which produces a WFE noise limit, $\sigma_n$, that can be compared with $\sigma_{10}$. 

\begin{figure}[h]
    \centering
    \includegraphics[width=0.7\linewidth]{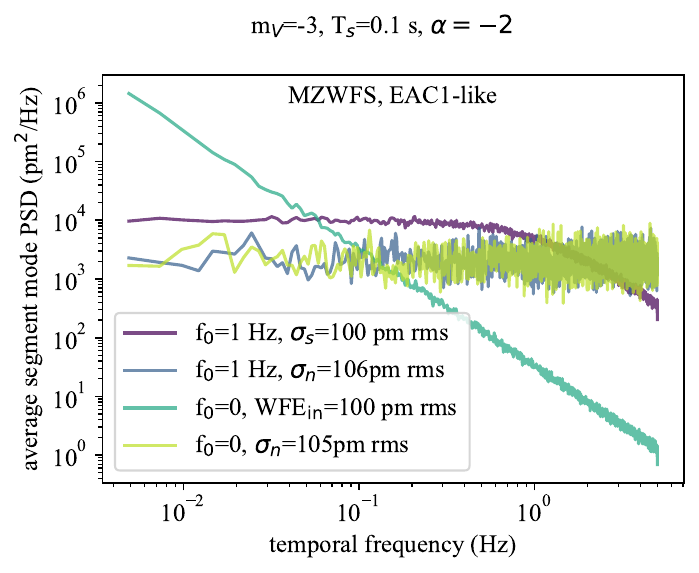}
    \caption{
    \red{
    The photon noise limits ($\sigma_{n}$) determined here are not dependent on any input power spectral density shape assumptions, demonstrated by the two different input PSDs ($\sigma_s$) that have different f$_0$ values but are normalized to the same input WFE produce effectively the same photon noise limits.
    }
    }
    \label{fig:f0_noise}
\end{figure}

Lastly, in \S\ref{sec:wfs} a pure $\alpha=-2$, $f_0$=0 power law is used. Although it is true that this type of input disturbance is the most optimal for AO as this is exactly the inverse of the AO error transfer function slope (which is also explored further in this paper in \S\ref{sec:control} and Appendix \S\ref{sec:gui_tutorial}), the reader may wonder whether or not a non-zero $f_0$ changes the corresponding $\sigma_n$ noise limits. Figure \ref{fig:f0_noise} addresses this with simluations of a MZWFS in an EAC1-like configuration,
showing that $\sigma_n$ is effectively the same for the case of two very different $f_0$ values but both with the same  100 pm rms input WFE ($\sigma_s$). This further reinforces the conclusions in \S\ref{sec:wfs} to be more universal and independent on assumed input disturbance spectral shape.
\subsection{Low Order NGS WFS simulations}
\label{sec:methods_lo}
Our assumed PL design parameters are listed in Table~\ref{tab:sim_pl_design}. We restrict the design to 4 ports to maximize signal in each port while being able to sense 3 orthogonal aberration modes. The ports are arranged with one in the center and the other three uniformly spaced around it at a spacing of 27.4$\mu$m. The core size is set such that the output ports are all single-moded over the wavelength range under consideration. This is determined by computing the port's V-number, $V = \frac{2\pi}{\lambda} r_\text{core} \sqrt{n_\text{core}^2 - n_\text{clad}^2}$ and ensuring the family of Bessel functions $J_n(x)$ admits only one zero less than $V$. Based on this, we select a core diameter of 3.54 $\mu$m. 

We simulate the PL at all wavelengths of interest (502.5-587.5 nm, with five evenly spaced single-wavelength simulations) using the \textit{lightbeam} Python package\cite{lightbeam}, which implements the beam-propagation method for solving for the propagation of light through weakly-guiding waveguides. This simulation provides an image of the PL output given an electric field at the entrance. For the achromatic case, we use this simulation at the central wavelength. Further, we simulate spectrally dispersing the output, as done in previous PL WFS experiments\cite{Lin2025}, by concatenating the images from each single-wavelength simulation. Due to variations in the PL's mode structure as a function of wavelength, aberration-dependent responses average out and cause a loss of information; this information is recovered by dispersing the output. We normalize each single-wavelength image such that it has 1/5 of the total flux, and add Poisson noise to the combined wavelength-multiplexed image to get the PL simulation output used here.

\begin{table}[]
    \centering
    \begin{tabular}{|c|c|}
        \hline
        Core diameter & 3.54 $\mu$m\\
        Core-to-core spacing & 27.4 $\mu$m\\
        Cladding diameter & 125.4 $\mu$m \\
        Taper length & 10000 $\mu$m \\
        Taper factor & 7.14 \\
        Numerical aperture & 0.117 \\
        Core refractive index & 1.4631\\
        Cladding refractive index & 1.4584 \\
        Jacket refractive index & 1.4530\\\hline
    \end{tabular}
    \caption{4-port simulated PL design parameters}
    \label{tab:sim_pl_design}
\end{table}

We simulate the optical system around the PL using the \textit{hcipy} Python package\cite{hcipy}. We simulate a 6m diameter telescope and inject light into the PL with an f-number of 5.656, chosen to maximize the throughput of the PL for a flat wavefront. We calibrate a linear wavefront reconstructor using the usual AO procedure, by measuring the differential image (normalized by dividing by the total flux through all ports summed together) when applying small positive and negative pokes and combining these to form an interaction matrix. This is then inverted using a Moore-Penrose pseudoinverse with a singular value cutoff of 1/30 to form a command matrix. The result of multiplying a reduced PL image (normalized, and with the signal for a flat wavefront subtracted off) by this matrix is the reconstructed wavefront. We confirmed that in the noiseless case that each mode is ideally reconstructed with no crosstalk.

\begin{figure}[h]
    \centering
    \includegraphics[width=1.0\linewidth]{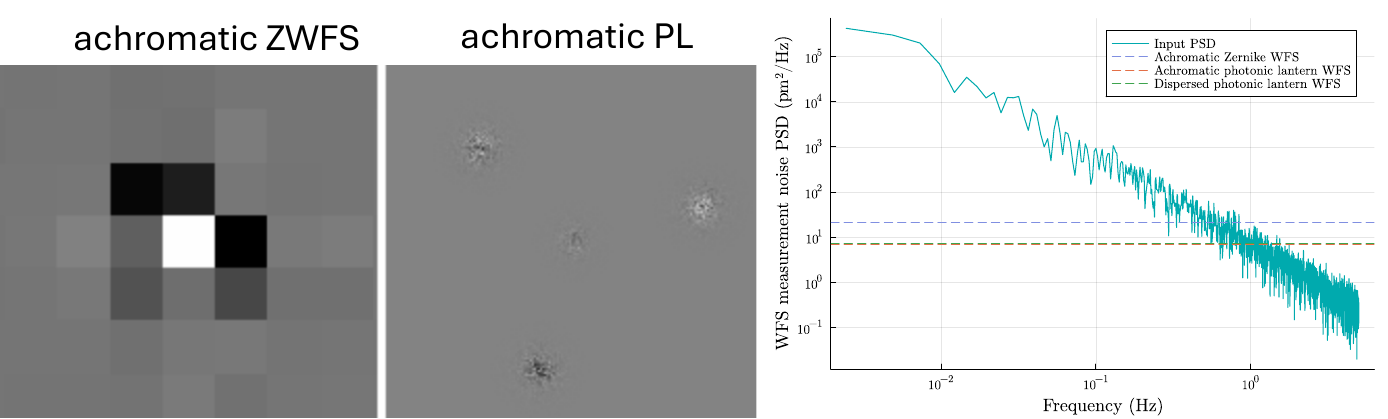}
    \caption{
    \red{
    Noisy target WFS images for a m$_V$=3 star for the achromatic ZWFS and PL (left and middle, respectively) that are the first of 6000 images used to reconstruct tip, tilt, and focus and average these corresponding PSDs  (right) to ultimately generate $\sigma_s$ and $\sigma_n$, showing the improved PL ($\sigma_n\lesssim8$ pm) vs. ZWFS ($\sigma_n\sim22$ pm) sensitivity.
    }
    }
    \label{fig:lowfs_illustration}
\end{figure}

We generate a representative open-loop time-series in each of tip, tilt, and focus via the identical process described in \S\ref{sec:wfs} and Appendix \ref{sec:methods_ho}.
For each time step, we compute the PL output and the reconstructed wavefront via a VMM as described above. We then compute the root-mean-square error (RMSE) of the difference between the noiseless reconstruction and the ground truth, which represent non-linearity errors. We find that this RMSE is 0.032pm in this case; this is determined by the resolution used in the simulation and is unlikely to impact this analysis. Figure \ref{fig:lowfs_illustration} illustrates the product of above-discussed steps and $\sigma_n$ is determined at $m_v$=3 (i.e., which generates one point of Fig. \ref{fig:measurement_noise} for each of the three curves in that figure).
\section{\red{1D GUI Tutorial}}
\label{sec:gui_tutorial}
\begin{figure}[!h]
    \centering
    \includegraphics[width=1.0\textwidth]{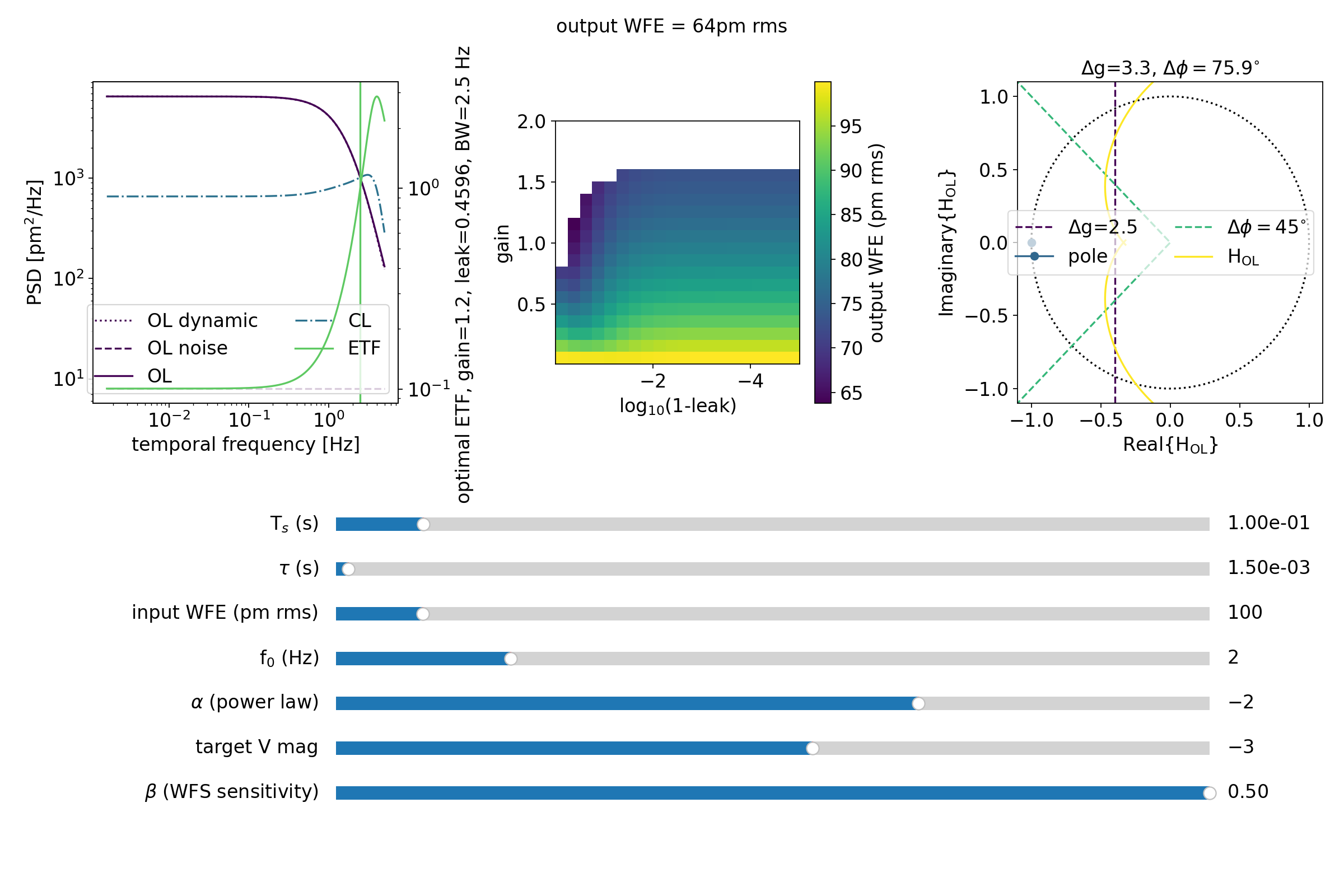}
    \caption{\red{A GUI snapshot, where in the shown instance the AO loop speed (1/T$_s$) of 10 Hz causes a 64 pm rms AO residual. The reader can adjust the T$_S$ and $m_V$ knobs while keeping the other parameters shown here unchanged to find that an minimum AO loop speed of 137 Hz is needed to reach $\sigma_{10}$. We encourage the reader to explore all adjustable dimensions of the GUI to understand how $\sigma_{10}$ requirements can be met or not.}}
    \label{fig:gui_tutorial}
\end{figure}

\red{
We encourage the reader to visit \url{https://github.com/LLNL/hwo1dGUI} and then click the mybinder.org link within the readme file that will open up a cloud-based web browser to run the GUI. Figure \ref{fig:gui_tutorial} illustrates one possible outputs of which there are many more for the reader to explore. 
}
\red{

The GUI title shows the integrated AO system closed-loop WFE between the WFS' Nyquist limit and 1/(10 minutes), which when equal to 10 pm rms is by definition $\sigma_{10}$. Adjustable variables include the WFS integration time (T$_s$; which sets the AO system frame rate to 1/T$_s$ by assuming a 100\% WFS camera duty cycle), AO system latency ($\tau$), input integrated WFE over the same frequency range which the titled output WFE is calculated, the high frequency power law $\alpha$, the spectral break (f$_0$), the LGS apparent magnitude ($m_V$) using the same flux and throughput calibration as Ref. \citenum{Douglas2019}, and WFS sensitivity ($\beta$)\cite{Guyon2005}.

The left panel in the GUI shows both (1) temporal PSD space for various open loop (OL) and closed-loop (CL) components on the left y-axis, and (2) AO error transfer function (ETF) space on the right y-axis. Open loop PSD components include (a) the noise component (labeled ``OL noise'' and adjusted by changing $m_V$ and/or $\beta$) and (b) the dynamic component (labeled "OL dynamic"), governed by the broken power law that is equation 2 in \citenum{Douglas2019}, whose shape is set by the spectral break ($f_0$) and whose high frequency slope is set by $\alpha$, and whose overall integrated normalization is set by the ``input WFE'' slider bar variable. In Ref. \citenum{Potier2022}, in general $f_0=2$ Hz and $\alpha=-2$, and in the worse case scenario for some spatial modes the input WFE is 100 pm rms, which motivates our chosen or noted use of these inputs in Figures \ref{fig:limits} and \ref{fig:gui_tutorial} and elsewhere throughout this paper. The labeled ``OL'' solid purple curve is a sum of the ``OL dynamic'' curves and ``OL noise'' curves and thus physically represents both the bandwidth and measurement error budget terms. The closed-loop temporal PSD (dashed-dotted blue line labeled ``CL'') is a product of the open loop PSD and the optimal ETF calculated and shown on right y axis. Our ETF model follows the standard Laplace domain AO transfer function model \cite{Alloin1994} that includes the WFS discrete sampling, system latency, the controller (for which we use a leaky integrator), and the DM zero order hold. The right y-axis of the left panel lists the optimal gain and leak parameters and the corresponding ETF 0 dB bandwidth (BW), which is also noted by the vertical green line in the plot.

The middle panel provides a more detailed view of how the optimal control gain and leak are calculated. The panel shows a 2D grid of gain and leak values for which at each value the output WFE is computed and illustrated by the heatmap and corresponding colorbar. The optimal gain and leak values are thus chosen based on the pair that produces the lowest output temporal WFE, similar to the process developed initially by Ref. \citenum{Gendron1994}. The whited out regions of the displayed gain and leak values are unstable controllers, determined by the Nyquist diagram in the right panel.

The right panel shows a Nyquist diagram--a standard method in control theory to assess stability\cite{Ogata1970}--for the complex-valued open loop AO transfer function (H$_{\text{OL}}$, both for positive and negative frequencies. The gain and phase margins ($\Delta$g and $\Delta\phi$, respectively), are computed with respect to the pole at (Imaginary\{H$_\text{OL}$\},Real\{H$_\text{OL}$\})=(-1,0), which represents a maximally unstable controller. We use the standard AO gain and phase margins\cite{Veran2009} of 2.5 and 45$^{\circ}$, illustrated in the plot and legend by the dashed purple and green lines, respectively. A gain and/or phase margin less than or equal to these values is considered unstable and removed from consideration in the middle panel (i.e., ``whited out''). Since the code automatically computes control parameters that are stable, the computed gain and phase margins (i.e., corresponding to the chosen gain and leak in the left panel right y-axis) in the title of this right panel will always meet these stability criteria.

}

\subsection* {Code, Data, and Materials Availability} 
Some code presented in this paper is available here: \url{https://github.com/LLNL/hwo1dGUI}.
Other Code, Data, and Materials underlying the results presented in this paper are not publicly available at this time but may be obtained from the authors upon reasonable request. 
\subsection* {Acknowledgments}
Components work were initially published in Ref. \citenum{Gerard2025}.

We thank Keith Jahoda, Aki Roberge, Ewan Douglas, Kerri Cahoy, Jared Males, and Olivier Guyon for discussions that led to the development of work presented in this paper. We also thank Garreth Ruanne, Jonathan Tesch, Breann Sitarski, Michael McElwain, Mitchell Troy, Alice Liu, Alden Jurling, and Tyler Groff for discussions that informed the content in this paper. 

This work was funded by the LLNL Laboratory Directed Research and Development (LDRD) Program (25-ERD-003). This document number is LLNL-JRNL-2013695. This work was performed under the auspices of the U.S.\ Department of Energy by Lawrence Livermore National Laboratory under Contract DE-AC52-07NA27344.


\bibliography{report}   
\bibliographystyle{spiejour}   


\vspace{2ex}\noindent\textbf{Benjamin L. Gerard} is a Research Scientist at Lawrence Livermore National Laboratory. He received his PhD degree in Physics and Astronomy from the University of Victoria in 2020 and was then a Postdoctoral Scholar at the University of California Santa Cruz before starting at LLNL in 2022. His current research interests include adaptive optics and exoplanet imaging. He is a member of SPIE.

\vspace{1ex}
\noindent Biographies and photographs of the other authors are not available.

\listoffigures
\listoftables

\end{spacing}
\end{document}